\documentclass[final,twocolumn,epjc3]{svjour3}          

\RequirePackage[T1]{fontenc}
\usepackage{multirow}
\smartqed  

\RequirePackage{graphicx}
\RequirePackage{mathptmx}      
\RequirePackage{flushend}
\RequirePackage[numbers,sort&compress]{natbib}
\RequirePackage[colorlinks,citecolor=blue,urlcolor=blue,linkcolor=blue]{hyperref}
\usepackage{amsmath}
\usepackage{amssymb}
\usepackage{epstopdf}

\journalname{Eur. Phys. J. C}

\begin{document}
\title{Dynamical quark loop light-by-light contribution to muon g-2 within the
nonlocal chiral quark model}
\author{
        A.~E.~Dorokhov\thanksref{eD,addrJINR,addrMSU}
        \and
        A.~E.~Radzhabov\thanksref{eR,addrIDSTU} 
        \and
        A.~S.~Zhevlakov\thanksref{eZ,addrIDSTU,addrTSU} 
}

\thankstext{eD}{e-mail: dorokhov@theor.jinr.ru}
\thankstext{eR}{e-mail: aradzh@icc.ru}
\thankstext{eZ}{e-mail: zhevlakov@phys.tsu.ru}

\institute{ Bogoliubov Laboratory of Theoretical Physics, JINR, 141980 Dubna, Russia\label{addrJINR}
          \and
          N.N.Bogoliubov Institute of Theoretical Problems of Microworld, M.V.Lomonosov Moscow State University, Moscow 119991, Russia\label{addrMSU}
          \and
          Institute for System Dynamics and Control Theory SB RAS, 664033 Irkutsk, Russia \label{addrIDSTU}
          \and
         Department of Physics, Tomsk State University, Lenin ave. 36, 634050 Tomsk, Russia \label{addrTSU}
}

\date{Received: date / Accepted: date}

\maketitle

\begin{abstract}
The hadronic corrections to the muon anomalous magnetic moment $a_{\mu}$, due
to the gauge-invariant set of diagrams with dynamical quark loop
light-by-light scattering insertions, are calculated in the framework of the
nonlocal chiral quark model. These results complete calculations of all
hadronic light-by-light scattering contributions to $a_{\mu}$ in the leading
order in the $1/N_{c}$ expansion. The result for the quark loop contribution
is $a_{\mu}^{\mathrm{HLbL,Loop}}=\left(  11.0\pm0.9\right)  \cdot10^{-10},$
and the total result is $a_{\mu}^{\mathrm{HLbL,N\chi QM}}=\left(
16.8\pm1.2\right)  \cdot10^{-10}$.
\end{abstract}

\section{Introduction}

Experimental and theoretical research on lepton anomalous magnetic moments has
a long and prominent history\footnote{For comprehensive reviews see
\cite{Jegerlehner:2009ry,Miller:2012opa,Dorokhov:2014iva,Knecht:2014sea}.}.
The most recent and precise measurements of the muon anomalous magnetic moment
$a_{\mu}$ were published in 2006 by the E821 collaboration at the Brookhaven
National Laboratory \cite{Bennett:2006fi}. The combined result, based on
nearly equal samples of positive and negative muons, is
\begin{equation}
a_{\mu}^{\mathrm{BNL}}=116~592~08.0~(6.3)\times10^{-10}\quad\lbrack
0.54~\mathrm{ppm}]. \label{amuBNL}%
\end{equation}
Later on, this value was corrected \cite{Mohr:2012tt,Agashe:2014kda} for a
small shift in the ratio of the magnetic moments of the muon and the proton
as
\begin{equation}
a_{\mu}^{\mathrm{BNL,CODATA}}=116~592~09.1~(6.3)\times10^{-10}.
\label{amuBNL2}%
\end{equation}
This exiting result is still limited by the statistical errors, and proposals
to measure $a_{\mu}$ with a fourfold improvement in accuracy were suggested at
Fermilab (USA) \cite{Venanzoni:2012sq} and J-PARC (Japan) \cite{Saito:2012zz}.
These plans are very important in view of a very accurate prediction of
$a_{\mu}$ within the standard model (SM). The dominant contribution in the SM
comes from QED
\begin{equation}
a_{\mu}^{\mathrm{QED}}=116~584~71.8951(80)\times10^{-10}\qquad
\text{\cite{Aoyama:2012wk}}. \label{aQED}%
\end{equation}
Other contributions are due to the electroweak corrections
\cite{Czarnecki:2002nt,Gnendiger:2013pva}%
\begin{equation}
a_{\mu}^{\mathrm{EW}}=15.36(0.1)\times10^{-10}\qquad
\text{\cite{Gnendiger:2013pva}}, \label{aEW}%
\end{equation}
the hadron vacuum polarization (HVP) contributions in the leading,
next-to-leading and next-next-to-leading order
\cite{Davier:2010nc,Hagiwara:2011af,Kurz:2014wya},
\begin{align}
a_{\mu}^{\mathrm{HVP,LO}}  &  =&694.91(3.72)(2.10)\times10^{-10}\qquad
\text{\cite{Hagiwara:2011af}},\label{aLO}\\
a_{\mu}^{\mathrm{HVP,NLO}}  &  =&-9.84(0.06)(0.04)\times10^{-10}\qquad
\text{\cite{Hagiwara:2011af}},\label{aNLO}\\
a_{\mu}^{\mathrm{HVP,NNLO}}  &  =&1.24(0.01)\times10^{-10}\qquad
\text{\cite{Kurz:2014wya}}, \label{aNNLO}%
\end{align}
and the hadronic light-by-light (HLbL) scattering contribution (as it is
estimated in \cite{Prades:2009tw})%
\begin{equation}
a_{\mu}^{\mathrm{HLbL}}=10.5(2.6)\cdot10^{-10}. \label{aHLbLprades}%
\end{equation}
As a result, the total value for the SM contribution, if we take
(\ref{aHLbLprades}) for HLbL, is
\begin{equation}
a_{\mu}^{\mathrm{SM}}=116~591~84.1~(5.0)\times10^{-10}. \label{aSMres}%
\end{equation}

From the comparison of (\ref{amuBNL2}) with (\ref{aSMres}) it follows that
there is a $3.11$ standard deviation between theory and experiment. This might
be an evidence for the existence of new interactions and stringently
constrains the parametric space of hypothetical interactions extending the SM
\footnote{In this regard we would like to mention the work
\cite{Queiroz:2014zfa}, where a public code for computing new physics
contributions to $a_{\mu}$ applicable to any particle physics models is
developed.}.

From above it is clear, that the main source of theoretical uncertainties
comes from the hadronic contributions. The HVP\ contribution $a_{\mu
}^{\mathrm{HVP,LO}}$, using analyticity and unitarity, can be expressed as a
convolution integral over the invariant mass of a known kinematical factor and
the total $e^{+}e^{-}\rightarrow\gamma^{\ast}\rightarrow$ hadrons
cross-section \cite{BM61}. Then the corresponding error in $a_{\mu
}^{\mathrm{HVP,LO}}$ essentially depends on the accuracy in the measurement of
the cross-section \cite{Davier:2010nc,Hagiwara:2011af}. In near future it is
expected, that new and precise measurements from CMD3 and SND at VEPP-2000 in
Novosibirsk, BES III in Beijing and KLOE-2 at DAFNE in Frascati will allow to
significantly increase the accuracy of the predictions for $a_{\mu
}^{\mathrm{HVP,LO}}$.

On the other hand, the HLbL contribution $a_{\mu}^{\mathrm{HLbL}}$ cannot be
calculated from first principles or (unlike to HVP) directly extracted from
phenomenological considerations. Instead, it has to be evaluated using various
QCD inspired hadronic models that correctly reproduce basic low- and high-
energy properties of the strong interaction. Nevertheless, as will be
discussed below, it is important for model calculations, that phenomenological
information and well established theoretical principles should significantly
reduce the number of model assumptions and the allowable space of model parameters.

Different approaches to the calculation of the contributions from the HLbL
scattering process to $a_{\mu}^{\mathrm{HLbL}}$ have been suggested. These
approaches can be classified into several types. The first one consists of
various extended versions of the vector meson dominance model (VMD)
supplemented by the ideas of the chiral effective theory, such as the hidden
local symmetry model (HLS) \cite{Hayakawa:1995ps}, the lowest meson dominance
(LMD) \cite{Knecht:2001qf,Melnikov:2003xd,Nyffeler:2009tw}, and the
(resonance) chiral perturbative theory ((R)$\chi$pT)
\cite{Kampf:2011ty,Roig:2014uja,Engel:2013kda}. The second type of approaches
is based on the consideration of effective models of QCD that use the
dynamical quarks as effective degrees of freedom. The rest include different
versions of the (extended) Nambu--Jona-Lasinio model (E)NJL
\cite{Bijnens:1995cc}, the constituent quark models with local interaction
(CQM)
\cite{Pivovarov:2001mw,Bartos:2001pg,Erler:2006vu,Boughezal:2011vw,Greynat:2012ww}%
, the models based on nonperturbative quark-gluon dynamics, like the nonlocal
chiral quark model (N$\chi$QM)
\cite{Dorokhov:2003kf,Dorokhov:2004ze,Dorokhov:2005pg,Dorokhov:2005ff,Dorokhov:2008pw,Dorokhov:2011zf,Dorokhov:2012qa}%
, the Dyson-Schwinger model \cite{Fischer:2010iz} (DS), or the holographic
models (HM) \cite{Hong:2009zw,Cappiello:2010uy}. More recently, there have
been attempts to estimate $a_{\mu}^{\mathrm{HLbL}}$ within the dispersive
approach (DA) \cite{Colangelo:2014dfa,Pauk:2014rfa} and the so-called rational
approximation (RA) approach \cite{Masjuan:2012qn}.

The aim of this work is to complete calculations of the leading in $1/N_{c}$
HLbL contributions within the N$\chi$QM started in
\cite{Dorokhov:2011zf,Dorokhov:2012qa} and compare the result with
(\ref{aHLbLprades}). Namely, in previous works we made detailed calculations
of hadronic contributions due to the exchange diagrams in the channels of
light pseudoscalar and scalar mesons. In the present work, the
detailed calculation of the light quark loop contribution is
given\footnote{Preliminary results of this work were announced in
\cite{Dorokhov:2014iva}.}.

\section{Light-by-light contribution to $a_{\mu}$ in the general case}

We start from some general consideration of the connection between the muon
AMM and the light-by-light (LbL) scattering polarization tensor. The muon AMM
for the LbL contribution can be extracted by using the projection
\cite{Aldins:1970id}
\begin{equation}
a_{\mu}^{\mathrm{LbL}}=\frac{1}{48m_{\mu}}\mathrm{Tr}\left(  (\hat{p}+m_{\mu
})[\gamma^{\rho},\gamma^{\sigma}](\hat{p}+m_{\mu})\mathrm{\Pi}_{\rho\sigma
}(p,p)\right)  , \label{aLbL}%
\end{equation}
where%
\begin{align}
&  \mathrm{\Pi}_{\rho\sigma}(p^{\prime},p)=e^{6}\int\frac{d^{4}q_{1}}%
{(2\pi)^{4}}\int\frac{d^{4}q_{2}}{(2\pi)^{4}}\frac{1}{q_{2}^{2}(q_{1}%
+q_{2})^{2}(q_{1}+k)^{2}}\times\label{Prs}\\
&  \quad\quad\times\gamma^{\mu}\frac{\hat{p}^{\prime}-\hat{q}_{2}+m_{\mu}%
}{(p^{\prime}-q_{2})^{2}-m_{\mu}^{2}}\gamma^{\nu}\frac{\hat{p}+\hat{q}%
_{1}+m_{\mu}}{(p+q_{1})^{2}-m_{\mu}^{2}}\gamma^{\lambda}\times\nonumber\\
&  \quad\quad\times\frac{\partial}{\partial k_{\rho}}\mathrm{\Pi}_{\mu
\nu\lambda\sigma}(q_{2},-(q_{1}+q_{2}),k+q_{1},-k),\nonumber
\end{align}
where $m_{\mu}$ is the muon mass, $k_{\mu}=(p^{\prime}-p)_{\mu},$ and it is
necessary to make the static limit $k_{\mu}\rightarrow0$ after
differentiation. Let us introduce the notation
\begin{align}
\frac{\partial}{\partial k_{\rho}}\mathrm{\Pi}_{\mu\nu\lambda\sigma}%
(q_{2},-(q_{1}+q_{2}),k+q_{1},-k)=\nonumber\\
{\Pi}_{\rho\mu\nu\lambda\sigma}%
(q_{2},-(q_{1}+q_{2}),q_{1})+O(k) \label{Pbold}%
\end{align}
for the derivative of the four-rank polarization tensor\footnote{First note,
the tensor $\mathrm{\Pi}_{\mu\nu\lambda\sigma}$ can be of any nature (QED,
hadronic, etc.) Another note concerns the important result expressing the
tensor ${\Pi}_{\rho\mu\nu\lambda\sigma}$ in the explicitly
gauge-invariant form that was obtained in \cite{Kuraev:1989cq,Arbuzov:2011zzc}.}, and rewrite
Eqs. (\ref{aLbL}) and (\ref{Prs}) in the form $\left(  q_{3}\equiv q_{1}%
+q_{2}\right)  $
\begin{align}
&a_{\mu}^{\mathrm{LbL}}=\frac{e^{6}}{48m_{\mu}}\int\frac{d^{4}q_{1}}{(2\pi
)^{4}}\int\frac{d^{4}q_{2}}{(2\pi)^{4}}\times\nonumber\\
&\quad\times\frac{{\Pi}_{\rho\mu\nu
\lambda\sigma}(q_{2},-q_{3},q_{1})\mathrm{T}^{\rho\mu\nu\lambda\sigma}\left(
q_{1},q_{2},p\right)  }{q_{1}^{2}q_{2}^{2}q_{3}^{2}((p+q_{1})^{2}-m_{\mu}%
^{2})((p-q_{2})^{2}-m_{\mu}^{2})}, \label{aLbL2}%
\end{align}
where the tensor $\mathrm{T}^{\rho\mu\nu\lambda\sigma}$ is the Dirac trace
\begin{align}
&\mathrm{T}^{\rho\mu\nu\lambda\sigma}\left(  q_{1},q_{2},p\right)
=
\mathrm{Tr}\biggl(  (\hat{p}+m_{\mu})[\gamma^{\rho},\gamma^{\sigma}](\hat
{p}+m_{\mu})\times\nonumber\\
&\quad\quad\times\gamma^{\mu}(\hat{p}-\hat{q}_{2}+m_{\mu})\gamma^{\nu}(\hat{p}%
+\hat{q}_{1}+m_{\mu})\gamma^{\lambda}\biggr)  .\nonumber
\end{align}

Taking the Dirac trace, the tensor $\mathrm{T}^{\rho\mu\nu\lambda\sigma}$
becomes a polynomial in the momenta $p$, $q_{1}$, $q_{2}$.
After that,
it is convenient to convert all momenta into the Euclidean space, and we will
use the capital letters $P$, $Q_{1}$, $Q_{2}$ for the corresponding
counterparts of the Minkowskian vectors $p$, $q_{1}$, $q_{2}$, e.g.
$P^{2}=-p^{2}=-m_{\mu}^{2}$, $Q_{1}^{2}=-q_{1}^{2}$, $Q_{2}^{2}=-q_{2}^{2}$.
Then Eq. (\ref{aLbL2}) becomes
\begin{align}
a_{\mu}^{\mathrm{LbL}}  &  =\frac{e^{6}}{48m_{\mu}}\int\frac{d_{E}^{4}Q_{1}%
}{(2\pi)^{4}}\int\frac{d_{E}^{4}Q_{2}}{(2\pi)^{4}}\frac{1}{Q_{1}^{2}Q_{2}%
^{2}Q_{3}^{2}}\frac{\mathrm{T}^{\rho\mu\nu\lambda\sigma}{\Pi}_{\rho
\mu\nu\lambda\sigma}}{D_{1}D_{2}},\nonumber\\
&  D_{1}=(P+Q_{1})^{2}+m_{\mu}^{2}=2(P\cdot Q_{1})+Q_{1}^{2},\label{aLbL3}x\\
&  D_{2}=(P-Q_{2})^{2}+m_{\mu}^{2}=-2(P\cdot Q_{2})+Q_{2}^{2}.\nonumber
\end{align}

Since the highest order of the power of the muon momentum $P$ in
$\mathrm{T}^{\rho\mu\nu\lambda\sigma}$ is two \footnote{The possible
combinations with momentum $P$ are
\begin{align*}
&  (P\cdot Q_{1})^{2}=(P\cdot Q_{1})(D_{1}-Q_{1}^{2})/2,\quad(P\cdot
Q_{2})^{2}=-(P\cdot Q_{2})(D_{2}-Q_{2}^{2})/2,\\
&  (P\cdot Q_{1})(P\cdot Q_{2})=-(D_{1}-Q_{1}^{2})(D_{2}-Q_{2}^{2})/4,\\
\quad &  (P\cdot Q_{1})=(D_{1}-Q_{1}^{2})/2,\quad(P\cdot Q_{2})=-(D_{2}%
-Q_{2}^{2})/2.
\end{align*}
} and ${\Pi}_{\rho\mu\nu\lambda\sigma}$ is independent of \ $P$,
the factors in the integrand of (\ref{aLbL3}) can be rewritten as
\begin{align}
\frac{\mathrm{T}^{\rho\mu\nu\lambda\sigma}{\Pi}_{\rho\mu\nu
\lambda\sigma}}{D_{1}D_{2}}=\sum\limits_{a=1}^{6}A_{a}\tilde{{\Pi}}%
_{a}, \label{TPbold}%
\end{align}
with the coefficients%
\begin{align}
&A_{1}=\frac{1}{D_{1}},\quad A_{2}=\frac{1}{D_{2}},\quad A_{3}=\frac{(P\cdot
Q_{2})}{D_{1}},\quad A_{4}=\frac{(P\cdot Q_{1})}{D_{2}},\nonumber\\
&\quad A_{5}=\frac
{1}{D_{1}D_{2}},\quad A_{6}=1,
\end{align}
where all $P$-dependence is included in the $A_{a}$ factors, while
$\tilde{{\Pi}}_{a}$ are $P$-independent.

Then, one can average over the direction of the muon momentum $P$ (as was
suggested in \cite{Jegerlehner:2009ry} for the pion-exchange contribution)
\begin{align}
&\int\frac{d_{E}^{4}Q_{1}}{(2\pi)^{4}}\int\frac{d_{E}^{4}Q_{2}}{(2\pi)^{4}%
}\frac{A_{a}}{Q_{1}^{2}Q_{2}^{2}Q_{3}^{2}}...=\nonumber\\
&\quad\frac{1}{2\pi^{2}}%
\int\limits_{0}^{\infty}dQ_{1}\int\limits_{0}^{\infty}dQ_{2}\int%
\limits_{-1}^{1}dt\,\sqrt{1-t^{2}}\frac{Q_{1}Q_{2}}{Q_{3}^{2}}\langle
A_{a}\rangle...,\label{Av} %
\end{align}
where the radial variables of integration $Q_{1}\equiv\left\vert
Q_{1}\right\vert $ and $Q_{2}\equiv\left\vert Q_{2}\right\vert $ and the
angular variable $t=(Q_{1}\cdot Q_{2})/\left(  \left\vert Q_{1}\right\vert
\left\vert Q_{2}\right\vert \right)  $ are introduced.
The averaged $A_{a}$ factors are \cite{Jegerlehner:2009ry}
\begin{align}
&  \left\langle A\right\rangle _{1}=\left\langle \frac{1}{\mathrm{D}_{1}%
}\right\rangle =\frac{R_{1}-1}{2m_{\mu}^{2}}\,,\quad\left\langle
A\right\rangle _{2}=\left\langle \frac{1}{\mathrm{D}_{2}}\right\rangle
=\frac{R_{2}-1}{2m_{\mu}^{2}}\,,\nonumber\\
&  \left\langle A\right\rangle _{3}=\left\langle \frac{(P\cdot Q_{2}%
)}{\mathrm{D}_{1}}\right\rangle =+(Q_{1}\cdot Q_{2})\frac{\left(
1-R_{1}\right)  ^{2}}{8m_{\mu}^{2}}\,\,,\label{Aa}\\
&  \left\langle A\right\rangle _{4}=\left\langle \frac{(P\cdot Q_{1}%
)}{\mathrm{D}_{2}}\right\rangle =-(Q_{1}\cdot Q_{2})\frac{\left(
1-R_{2}\right)  ^{2}}{8m_{\mu}^{2}}\,\,,\,\,\,\nonumber\\
&  \left\langle A\right\rangle _{5}=\left\langle \frac{1}{\mathrm{D}%
_{1}\mathrm{D}_{2}}\right\rangle =\frac{1}{m_{\mu}^{2}Q_{1}Q_{2}x}%
\arctan\left[  \frac{zx}{1-zt}\right]  \,,\nonumber\\
&  \left\langle A\right\rangle _{6}=\left\langle 1\right\rangle =1,\nonumber
\end{align}
with%
\begin{align}
&  \quad x=\sqrt{1-t^{2}}\,,\quad R_{i}=\sqrt{1+\frac{4m_{\mu}^{2}}{Q_{i}^{2}%
}}\,~(i=1,2),\\
&  \quad z=\frac{Q_{1}Q_{2}}{4m_{\mu}^{2}}\left(  1-R_{1}\right)  \left(
1-R_{2}\right)  .\nonumber
\end{align}
After averaging
the LbL contribution can be represented in the form
\begin{equation}
a_{\mu}^{\mathrm{LbL}}=\int\limits_{0}^{\infty}dQ_{1}\int\limits_{0}^{\infty
}dQ_{2}\quad\rho^{\mathrm{LbL}}(Q_{1},Q_{2}), \label{aLbL4}%
\end{equation}
with the density $\rho^{\mathrm{LbL}}(Q_{1},Q_{2})$ being defined as
\begin{equation}
\rho^{\mathrm{LbL}}(Q_{1},Q_{2})=\frac{Q_{1}Q_{2}}{2\pi^{2}}\sum
\limits_{a=1}^{6}\int\limits_{-1}^{1}dt\,\frac{\sqrt{1-t^{2}}}{Q_{3}^{2}%
}\langle A_{a}\rangle\tilde{{\Pi}}_{a}. \label{dens}%
\end{equation}

Thus, the number of momentum integrations in the original expression for
(\ref{aLbL}) is reduced from eight to three. The transformations leading from
(\ref{aLbL}) to (\ref{aLbL4}), are of general nature, independent of the
theoretical (model) assumptions on the form of the polarization tensors
$\tilde{{\Pi}}_{a}$. In particular, this 3D-representation is common
for all hadronic LbL contributions: the pseudoscalar meson exchange
contributions \cite{Jegerlehner:2009ry,Dorokhov:2011zf}, the scalar meson
exchange contributions \cite{Dorokhov:2012qa}, and the quark loop
contributions discussed in the present work. The next problem to be elaborated
is the calculation of the density $\rho^{\mathrm{HLbL}}(Q_{1},Q_{2})$ in the framework of
the model.

\section{Hadronic Light-by-light contribution to $a_{\mu}$ within N$\chi$QM}

Let us briefly review the basic facts about the N$\chi$QM\footnote{More
detailed information about the model is contained in our previous works
\cite{Dorokhov:2008pw,Dorokhov:2012qa}.}. The Lagrangian of the $SU(3)$
nonlocal chiral quark model with $SU(3)\times SU(3)$ symmetry has the form
\begin{align}
\mathcal{L}  &  =\bar{q}(x)(i\hat{\partial}-m_{c})q(x)+\frac{G}{2}[J_{S}%
^{a}(x)J_{S}^{a}(x)+J_{PS}^{a}(x)J_{PS}^{a}(x)]\nonumber\\
&  -\frac{H}{4}T_{abc}\Big[J_{S}^{a}(x)J_{S}^{b}(x)J_{S}^{c}(x)-3J_{S}%
^{a}(x)J_{PS}^{b}(x)J_{PS}^{c}(x)\Big],\label{Model}
\end{align}
where $q\left(  x\right)  $ are the quark fields, $m_{c}$ $\left(  m_{c,u}%
=m_{c,d}\neq m_{c,s}\right)  $ is the diagonal matrix of the quark current masses,
and $G$ and $H$ are the four- and six-quark coupling constants.
The nonlocal structure of the model is introduced via the nonlocal quark
currents
\begin{equation}
J_{M}^{a}(x)=\int d^{4}x_{1}d^{4}x_{2}\,F(x_{1},x_{2})\,\bar{q}(x-x_{1}%
)\,\Gamma_{M}^{a}q(x+x_{2}), \label{JaM}%
\end{equation}
where $M=S$ for the scalar and $M=PS$ for the pseudoscalar channels,
$\Gamma_{{S}}^{a}=\lambda^{a}$, $\Gamma_{{PS}}^{a}=i\gamma^{5}\lambda^{a},$
and $F(x_{1},x_{2})$ is the form factor with the nonlocality parameter
$\Lambda$ reflecting the nonlocal properties of the QCD vacuum. The $SU(2)$
version of the N$\chi$QM with $SU(2)\times SU(2)$ symmetry is obtained by
setting $H$ to zero and taking only scalar-isoscalar and
pseudoscalar-isovector currents.

{ Within the N$\chi$QM, the standard mechanism for spontaneous breaking of
chiral symmetry occurs, which is typical for the Nambu--Jona-Lasinio type models with the chiral symmetric four-fermion interaction (local or nonlocal).
Due to this interaction the massless quark becomes massive, and in the hadron spectrum the gap between the massless (in the chiral limit) Nambu-Goldstone pion and the massive scalar meson appears. This feature is common for the models used for the calculation of the hadronic contributions to the muon $g-2$: the extended NJL model \cite{Bijnens:1995cc}, the constituent chiral quark model \cite{Greynat:2012ww}, the Dyson-Schwinger model \cite{Fischer:2010iz}, the nonlocal chiral quark model \cite{Dorokhov:2003kf,Dorokhov:2004ze,Dorokhov:2005pg,Dorokhov:2005ff,Dorokhov:2008pw,Dorokhov:2011zf,Dorokhov:2012qa}. In the nonlocal models the dynamically generated quark mass becomes momentum dependent and the inverse dynamical quark propagator takes the form}%
\begin{equation}
S^{-1}\left(  k\right)  =\widehat{k}-m(k^{2}) \label{S-1}%
\end{equation}
where $m(k^{2})=m_{c}+m_{D}F(k^{2},k^{2})$ is the dynamical quark mass
obtained by solving the Dyson-Schwinger equation.
{The significant feature of the nonlocal models \cite{Dorokhov:2003kf,Dorokhov:2004ze,Dorokhov:2005pg} is that they correctly interpolate between the low-energy region (and consistent with the
{
low-energy theorems}) and the high-energy region (where they are consistent with OPE). The basic fact is that the momentum dependent dynamical quark mass, that is the constituent quark mass $m(0)=m_{c}+m_{D}$ at low virtualities, becomes the current quark mass $m_c$ at large virtualities. This is in contrast to the local models, where the quark mass is the constituent one at any virtuality.}

For numerical estimates two
versions of the form factor (in momentum space) are used: the Gaussian form
factor
\begin{equation}
F_{G}\left(  k_{E}^{2},k_{E}^{2}\right)  =\mathrm{exp}\left(  -k_{E}%
^{2}/\Lambda^{2}\right)  , \label{fk}%
\end{equation}
and the Lorentzian form factor
\begin{equation}
F_{L}\left(  k_{E}^{2},k_{E}^{2}\right)  =\frac{1}{\left(  1+k_{E}^{2}%
/\Lambda^{2}\right)  ^{2}}. \label{fLorentzian}%
\end{equation}
The second version is used in order to test the stability of the results to
the nonlocality shape.

Next,
 it is necessary to introduce in the nonlocal chiral Lagrangian (\ref{Model}) the gauge-invariant interaction with an external
photon field $A_{\mu}(z)$ .
\begin{figure}[htb]
\begin{center}
\centerline{\begin{tabular*}{\columnwidth}{@{\extracolsep{\fill}}cccc}
\resizebox{!}{0.12\textheight}{\includegraphics{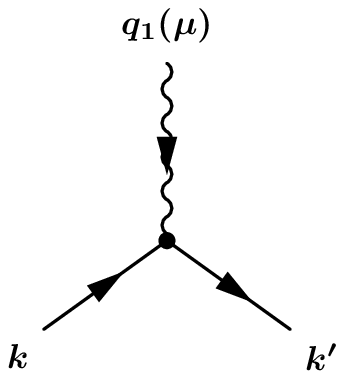}}
& \resizebox{!}{0.12\textheight}{\includegraphics{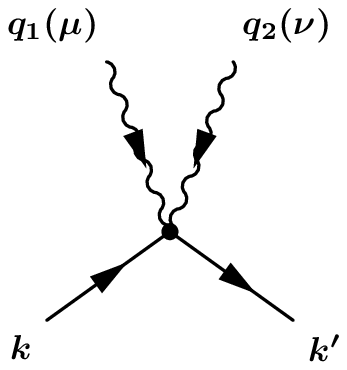}}\\
(a) & (b) \\\\ 
 \resizebox{!}{0.12\textheight}{\includegraphics{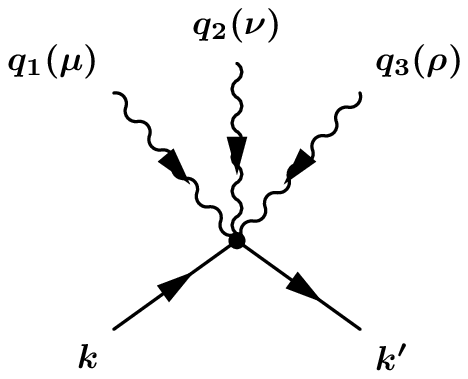}}
& \resizebox{!}{0.12\textheight}{\includegraphics{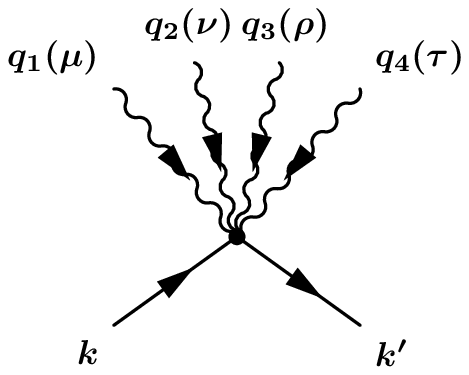}}
\\
 (c) & (d)
\end{tabular*}}
\end{center}
\caption{The quark-photon vertex $\mathrm{\Gamma}_{\mu}^{\left(  1\right)
}\left(  q\right)  $, the quark-two-photon vertex $\mathrm{\Gamma}_{\mu\nu
}^{(2)}\left(  q_{1},q_{2}\right)  $, the quark-three-photon vertex
$\Gamma_{\mu\nu\rho}^{(3)}(q_{1},q_{2},q_{3}),$ and the quark-four-photon
vertex $\Gamma_{\mu\nu\rho\tau}^{(4)}(q_{1},q_{2},q_{3},q_{4})$ . }%
\label{fig:VerticesPhot-n}%
\end{figure}
This can be made through the introduction of the path-ordered
Schwinger phase factor for the quark field as%
\begin{equation}
q\left(  y\right)  \rightarrow Q\left(  x,y\right)  =\mathcal{P}\exp\left\{
i\int_{x}^{y}dz^{\mu}A_{\mu}\left(  z\right)  \right\}  q\left(  y\right)  .
\label{SchwPhF}%
\end{equation}
Then, apart from the kinetic term, the additional, nonlocal terms in the
interaction of quarks with the gauge field are generated via substitution
\begin{align}
J_{M}^{a}(x)  &  \rightarrow J_{M}^{a}(x)=\int d^{4}x_{1}d^{4}x_{2}%
\,f(x_{1})f(x_{2})\times\nonumber\\
&  \,\times\,\overline{Q}(x-x_{1},x)\,\Gamma_{M}^{a}Q(x,x+x_{2}),
\label{JaMgauge}%
\end{align}
inducing the quark-antiquark--$n$-photon vertices.
In order to obtain the explicit form of these vertices, it is necessary to fix
the rules for the contour integral in the phase factor.
The 
scheme, based on the rules that the
derivative of the contour integral does not depend on the path shape
\begin{align}
   \frac{\partial}{\partial y^{\mu }}\int\limits_{x}^{y}dz^{\nu }\
   F_{\nu}(z)=F_{\mu }(y),\quad \delta^{(4)}\left( x-y\right)
   \int\limits_{x}^{y}dz^{\nu}\ F_{\nu }(z)=0,
\nonumber
\end{align}
was suggested in \cite{Mandelstam:1962mi}
and applied to nonlocal models in \cite{Terning:1991yt}.
For our purpose,  we need to consider the
quark-antiquark vertices with one-, two-, three- and four- photons (Fig. \ref{fig:VerticesPhot-n}).
The first two types of vertices was derived in \cite{Terning:1991yt}, the vertex with three photons was obtained in \cite{Liao:2008dd},
and the quark-4-photon vertex is given in the present work. Their explicit form and the definition for the finite-difference
derivatives $m^{(n)}(k,k')$ are presented in the Appendix.
The simplest quark-photon vertex has the usual local part as well as the nonlocal piece in terms of the first finite-difference derivative  $m^{(1)}(k,k')$
\begin{align}
&  \Gamma_{\mu}^{\left(  1\right)  }\left(  q_{1}\right)  =\gamma_{\mu}%
+\Delta\Gamma_{\mu}^{\left(  1\right)  }\left(  q_{1}\right)  , \label{GamTot}%
\\
&  \Delta\Gamma_{\mu}^{\left(  1\right)  }\left(  q_{1}\right)  =-\left(
k+k^{\prime}\right)  _{\mu}m^{\left(  1\right)  }\left(  k,k^{\prime}\right)
, \label{DGam1}%
\end{align}
while the quark-antiquark vertices with more than one photon insertion are purely
nonlocal (see Appendix).

\begin{figure*}[tb]
\centerline{\includegraphics[width=0.85\textwidth]{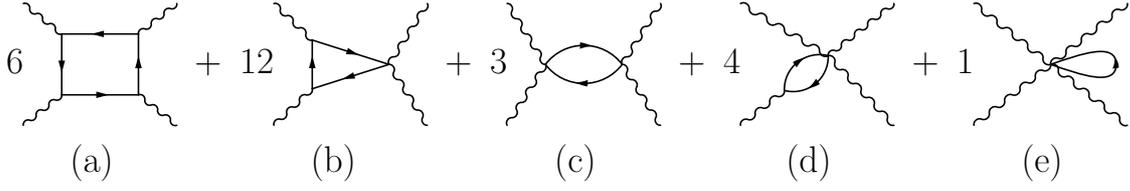}} \vspace*{8pt}%
\caption{The box diagram and the diagrams with nonlocal multiphoton
interaction vertices represent the gauge invariant set of diagrams contributing to the
polarization tensor $\mathrm{\Pi}_{\mu
\nu\lambda\rho}(q_{1},q_{2},q_{3},q_{4})$. The numbers in front of the diagrams are
the combinatoric factors. }%
\label{Fig: BoxCont}%
\end{figure*}

{
We have to remind, that in the models with the chiral symmetric four-quark interaction (nonlocal or local NJL type) the Goldstone particles and other mesons appear as the poles in the quark-antiquark scattering matrix due to the summation of infinite number of diagrams \cite{Bijnens:1995cc,Hayakawa:1995ps,Greynat:2012ww,Dorokhov:2008pw,Dorokhov:2011zf,Fischer:2010iz}. In these diagrams, the quark and antiquark interact via the four-quark interaction. On the other hand, in the box diagram (Fig. \ref{Fig: BoxCont}), the quark and antiquark do not interact between each other and thus it is separated from the set of diagrams producing mesons as bound states. It means in particular, that in these approaches there are no double-counting effects.
On the other hand in the framework NJL model it was shown that these two
type of contributions, e.g. box ans bound state one, are necessary for the correct description of such processes as
pion polarizability \cite{Dorokhov:1997rv} or $\pi\pi$-scattering \cite{Osipov:2007zz}
and omitting one of these contribution will lead to
large breaking of chiral symmetry.
}

{To above we can add that from the quark-hadron duality arguments, the quark loop (as for the two-point correlator as well for the four-point correlator) represents the contribution of the continuum of excited hadronic states. In the language of the spectral densities, the model calculations correspond to the model of the spectral density saturated by the lowest hadronic resonance plus the excited hadronic state continuum. The first part is for the meson-exchange diagrams, and the latter for the quark loop. It is the quark loop (continuum) provides the correct large photon momentum QCD asymptotics for the Adler function, three- and four- point correlators.}

With the Feynman rules for the dynamical quark propagator (\ref{S-1}) and the
quark-photon vertices (\ref{GamTot}), (\ref{DGam2}), (\ref{Gammappqqq}), and
(\ref{Gqqqq4}), the gauge invariant set of diagrams describing the
polarization tensor $\mathrm{\Pi}_{\mu\nu\lambda\sigma}(q_{2},-(q_{1}%
+q_{2}),k+q_{1},-k)$ due to the dynamical quark loop contribution is given in Fig. \ref{Fig: BoxCont}.

\section{The results}

For the numerical estimates, the $SU(2)$- and $SU(3)$- versions of the N$\chi
$QM model are used. In order to check the model dependence of the final
results, we also perform calculations for different sets of model parameters.

In the $SU(2)$ model, the same scheme of fixing the model parameters as in
\cite{Dorokhov:2011zf,Dorokhov:2012qa} is applied: fitting the parameters
$\Lambda$ and $m_{c}$ by the physical values of the $\pi^{0}$ mass and the
$\pi^{0}\rightarrow\gamma\gamma$ decay width, and varying $m_{D}$ in the
region 
 $150-400$ MeV. For estimation of  $a_{\mu}^{\mathrm{HLbL}}$ and its error, we use the region
for $m_{D}$ from $200$ to $350$ MeV. 

\begin{figure}[t]
\centerline{\includegraphics[width=0.45\textwidth]{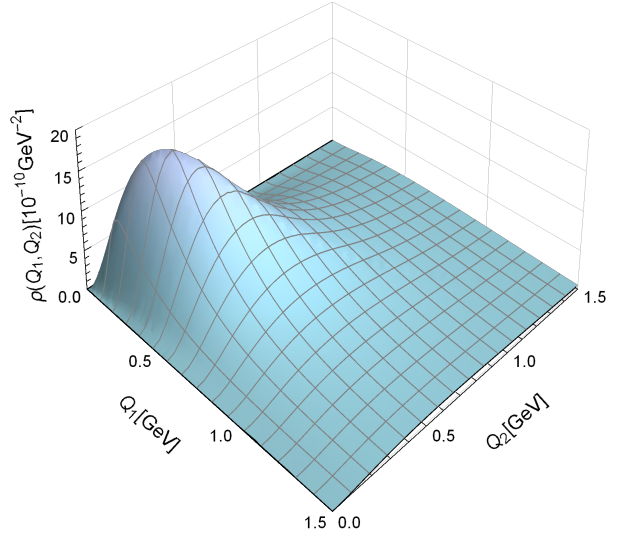}} \vspace*{8pt}%
\caption{The 3D density $\rho(Q_{1},Q_{2})$ defined in Eqs. (\ref{aLbL4},\ref{dens}).
}%
\label{Fig: 3Ddensity}%
\end{figure}

For the $SU(3)$ version of the model, it is necessary to fix two more
parameters: the current and dynamical masses of the strange quark. We suggest
to fix them by fitting the $K^{0}$ mass and obtaining more or less reasonable
values for the $\eta$ meson mass and the $\eta\rightarrow\gamma\gamma$ decay
width. The main problem here is that the lowest value for the nonstrange
dynamical mass $m_{D}$ is $240$ MeV, because at lower $m_{D}$ the $\eta$ meson
becomes unstable within the model approach.

Additionally, in order to show that the different schemes of parameter fixing will
lead to similar results for $a_{\mu}^{\mathrm{HLbL}}$, we calculate this quantity
for the model (\ref{Model}) with parameters taken from \cite{Scarpettini:2003fj}
for the Gaussian ($G_{I}-G_{IV}$) and the Lorentzian ($L_{I}-L_{IV}$) nonlocal form factors. The authors of \cite{Scarpettini:2003fj}
have used other scheme of parameter fixing. Namely,
the value of light current quark mass is fixed
($8.5$ MeV for $G_{I}-G_{III}$, $7.5$ MeV for $G_{IV}$, $4.0$ MeV for $L_{I}-G_{III}$, and $3.5$ MeV for $L_{IV}$).
The other parameters are fitted in order to reproduce the values
of the pion and kaon masses, the pion decay constant $f_\pi$, and, alternatively, the $\eta^\prime$ mass
for sets $G_{I}$, $G_{IV}$, $L_{I}$, $L_{IV}$ or the $\eta^\prime\rightarrow\gamma\gamma$ decay
width for sets $G_{II}$, $G_{III}$, $L_{II}$, $L_{III}$.

The important result, independent of the parameterizations, is the behavior of
the density $\rho^{\mathrm{HLbL}}(Q_{1},Q_{2}),$ shown in Fig.
\ref{Fig: 3Ddensity}. One can see, that $\rho^{\mathrm{HLbL}}(Q_{1},Q_{2})$ is
zero at the edges $\left(  Q_{1}=0~\mathrm{or}~Q_{2}=0\right)  $ and is
concentrated in the low-energy region\footnote{One should point out that the
density for the mesonic exchanges has similar behavior.} $\left(  Q_{1}\approx
Q_{2}\approx300~\mathrm{MeV}\right)  $ providing the dominant contribution to
$a_{\mu}^{\mathrm{HLbL}}$. This behavior at the edges appears to be due to
cancelations of contributions from different diagrams of Fig.
\ref{Fig: BoxCont}.

\begin{figure}[t]
\centerline{\includegraphics[width=0.45\textwidth]{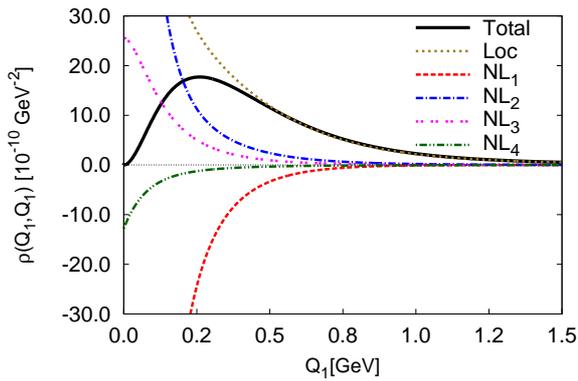}}
\vspace*{8pt}\caption{The 2D slice of the density $\rho(Q_{1},Q_{2})$ at
$Q_{2}=Q_{1}$. Different curves correspond to the contributions of
topologically different sets of diagrams drawn in Fig. \ref{Fig: BoxCont}. The
contribution of the box diagram with the local vertices, Fig.
\ref{Fig: BoxCont}a, is the dot (olive) line(Loc); the box diagram, Fig.
\ref{Fig: BoxCont}a, with the nonlocal parts of the vertices is the dash (red)
line (NL$_{1}$); the triangle, Fig. \ref{Fig: BoxCont}b, and loop, Fig.
\ref{Fig: BoxCont}c, diagrams with the two-photon vertices is the dash-dot
(blue) line (NL$_{2}$); the loop with the three-photon vertex, Fig.
\ref{Fig: BoxCont}d, is the dot-dot (magenta) line (NL$_{3}$); the loop with
the four-photon vertex, Fig. \ref{Fig: BoxCont}e, is the dash-dot-dot (green)
line (NL$_{4}$); the sum of all contributions (Total) is the solid (black)
line. At zero all contributions are finite.
}%
\label{Fig: Slice}%
\end{figure}

In Fig. \ref{Fig: Slice}, the slice of $\rho^{\mathrm{HLbL}}(Q_{1},Q_{2})$ in
the diagonal direction $Q_{2}=Q_{1}$ is presented together with the partial
contributions from the diagrams of different topology. One can see, that the
$\rho^{\mathrm{HLbL}}(0,0)=0$ is due to a nontrivial cancelation of different
diagrams of Fig. \ref{Fig: BoxCont}. This important result is a consequence of
gauge invariance and the spontaneous violation of the chiral symmetry, and
represents the low energy theorem analogous to the theorem for the Adler
function at zero momentum. Another interesting feature is, that the large
$Q_{1}$, $Q_{2}$ behavior is dominated by the box diagram with local vertices
and quark propagators with momentum-independent masses in accordance with perturbative theory.
All this is very important characteristics of the N$\chi$QM, interpolating the
well-known results of the chiral perturbative theory at low momenta and the
operator product expansion at large momenta. Earlier, similar results were
obtained for the two-point \cite{Dorokhov:2003kf,Dorokhov:2004ze} and
three-point \cite{Dorokhov:2005pg} correlators.

\begin{figure*}[t]
\begin{center}
\centerline{\includegraphics[width=0.45\textwidth]{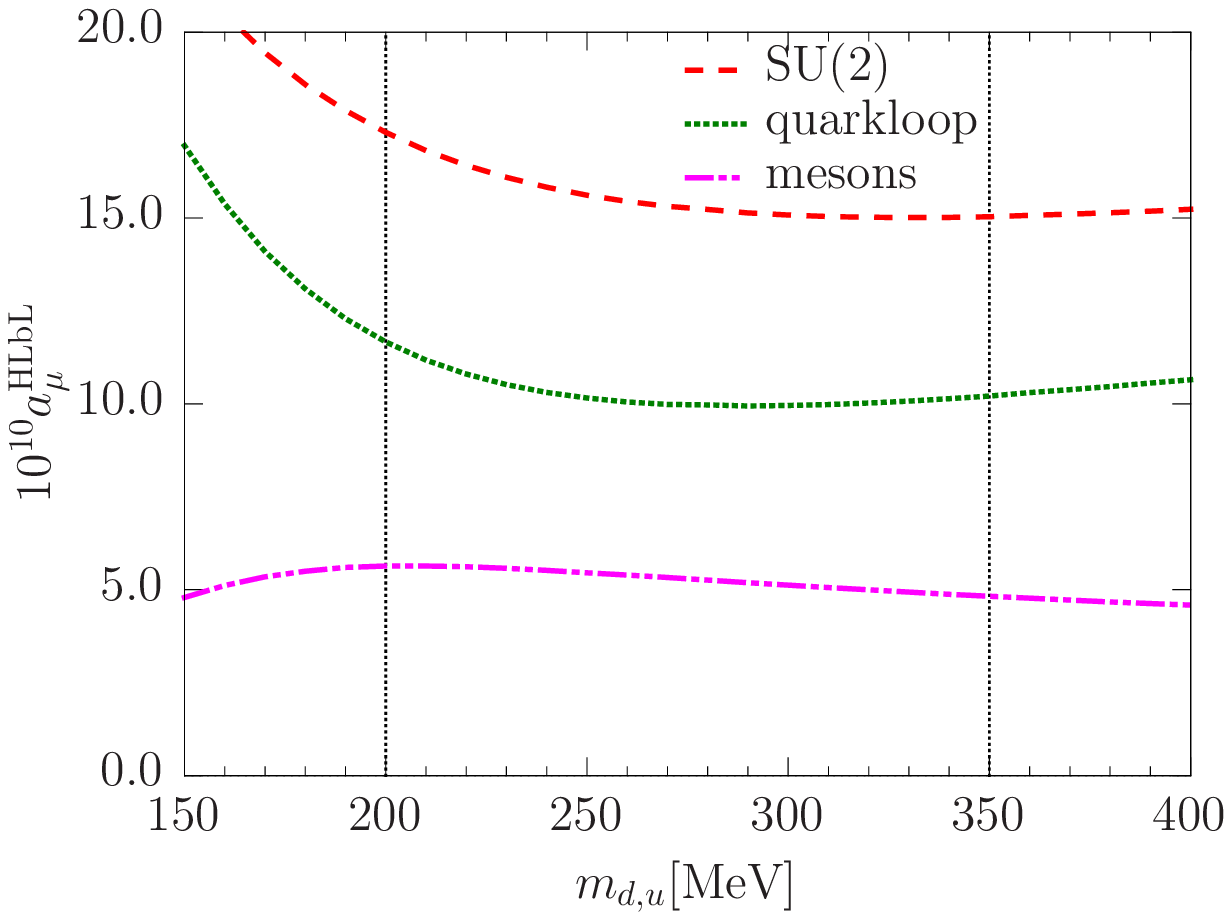} 
\includegraphics[width=0.45\textwidth]{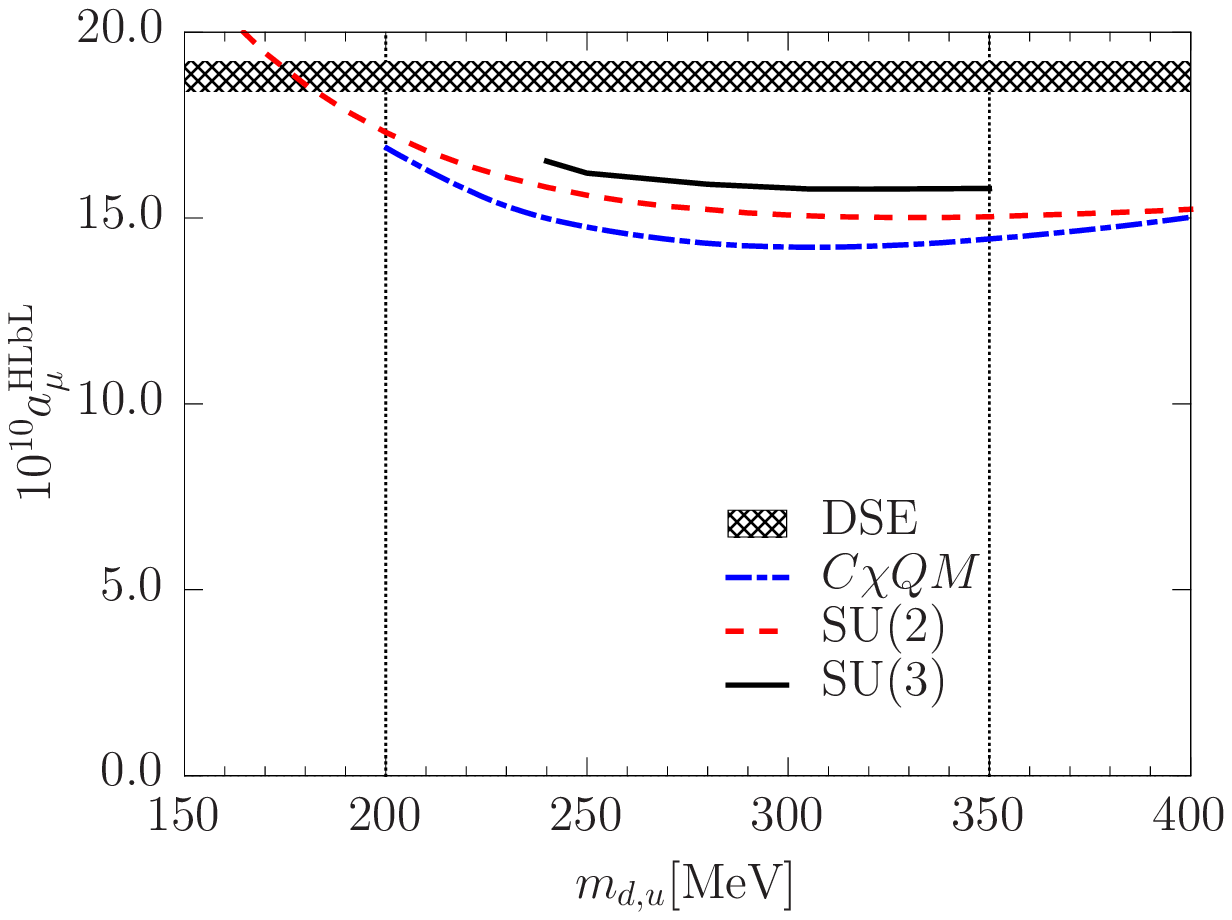}} 
\end{center}
\caption{(Left) The results for $a_{\mu}^{\mathrm{HLbL}}$ in the $SU(2)$ model: the
red dashed line is the total result, the green dotted line is the quark loop
contribution and the magenta dash-dot-dot line is the $\pi+\sigma$ contribution. Thin
vertical line indicates the region for estimation of $a_{\mu}^{\mathrm{HLbL}}$ error band.
(Right) The results for $a_{\mu}^{\mathrm{HLbL}}$: the black solid line is the $SU(3)$-result,
the red dash line corresponds to the $SU(2)$-result,
the blue dash-dotted line is the $C\chi QM$ result \cite{Greynat:2012ww},
hatched region correspond to DSE result \cite{Fischer:2010iz}.
}%
\label{fig: 16}%
\end{figure*}

The numerical results for the value of $a_{\mu}^{\mathrm{HLbL}}$ are given in
the table and presented
in Fig. 
\ref{fig: 16} for the $SU(2)$ and $SU(3)$ models
together with the result of
C$\chi $QM \cite{Greynat:2012ww} and DSE \cite{Fischer:2010iz}
calculations.
The estimates for the partial contributions to $a_{\mu}^{\mathrm{HLbL}}$ (in
$10^{-10}$) are the $\pi^{0}$ contribution $5.01(0.37)$ \cite{Dorokhov:2011zf}%
, the sum of the contributions from $\pi^{0}$, $\eta$ and $\eta^{\prime}$
mesons $5.85(0.87)$ \cite{Dorokhov:2011zf}, the scalar $\sigma$, $a_{0}(980)$
and $f_{0}(980)$ mesons contribution $0.34(0.48)$
\cite{Dorokhov:2014iva,Dorokhov:2012qa}, and the quark loop contribution is
$11.0(0.9)$ \cite{Dorokhov:2014iva}.
In all cases we estimate the absolute value of the result and its error by
calculating  $a_{\mu}^{\mathrm{HLbL,N\chi QM}}$ for the space of model parameters fixed by above mentioned
observables, except one, varying $m_D$. Because in all cases the resulting curves (Fig. \ref{fig: 16}) are quite smooth,
it gives to us a credit to point out rather small model errors ($\leq 10\%$) for the intermediate and final results.
Thus our claim is that the total contribution obtained in the
leading order in the $1/N_{c}$ expansion within the nonlocal chiral quark model is (see also \cite{Dorokhov:2014iva}%
)
\begin{equation}
a_{\mu}^{\mathrm{HLbL,N\chi QM}}=16.8(1.25)\cdot10^{-10}. \label{aNchiQM}%
\end{equation}
This value accounts for the spread of the results depending on reasonable variation of the model parameters and sensitivity to the different choice of the nonlocality shapes. Note, that as it was emphasized in \cite{Greynat:2012ww}, the results of these kind of calculations do not include the ''systematic error'' of the models.

Comparing with other model calculations, we conclude that
our results are quite close to the recent results obtained in
\cite{Fischer:2010iz,Greynat:2012ww}.\footnote{
In earlier works \cite{Bijnens:1995cc,Hayakawa:1995ps}, the quark loop contribution was found one order less than in more recent calculations. To our opinion, one of the reason for that, is that in those models the photon-quark coupling is suppressed by the VMD form factors.
}
{It is no accidental. The most close to our model is the Dyson-Schwinger model used in \cite{Fischer:2010iz}. The specific feature of both models is that the
{
kernel of } nonlocal
interaction is motivated by QCD. In \cite{Fischer:2010iz} the kernel of the interaction is generated by the nonperturbative gluon exchanges. In the N$\chi$QM the form of the kernel is motivated by the instanton vacuum models. The other difference between the N$\chi$QM and \cite{Fischer:2010iz} is that, in a sense, the N$\chi$QM has a minimal structure (with respect to number of Lorentz structures for vertices, etc). Nevertheless, the predictions of the N$\chi$QM for the different contributions to the muon $g-2$ are in agreement with \cite{Fischer:2010iz} within $10\%$.}

{ The constituent chiral quark model used in \cite{Greynat:2012ww} corresponds to the local limit of the N$\chi$QM. This limit is achieved when the nonlocality parameter $\Lambda$ goes to infinity, that means that the nonlocal form factors become constants: $F(k^2,p^2)\rightarrow1$. Taking this limit the N$\chi$QM becomes one-parametric one (only $M_q$) and we reproduce the $M_q$ dependence of quark box contribution to $a_\mu^{\mathrm{HLbL}}$ shown in Fig. 13 of \cite{Greynat:2012ww}. What is more interesting and important is that, the $M_q$ dependencies of total contribution to $a_\mu^{\mathrm{HLbL}}$ in \cite{Greynat:2012ww} (Fig. 14) and in the N$\chi$QM have the same qualitative behavior and very close (with less than $10\%$) qualitatively. This is clear from Fig. \ref{fig: 16}.}

{ These facts are very pleasant for the phenomenology of the HLbL contributions to the muon, because it means that even starting from the models that differ in many details, the predictions are still very stable numerically.}

\section{Conclusions}

In this paper, we have presented the results for the contribution of the
dynamical quark loop mechanism for the light-by-light scattering to the muon
anomalous magnetic moment within the nonlocal chiral quark model. In previous
works \cite{Dorokhov:2011zf,Dorokhov:2012qa,Dorokhov:2014iva}, we calculated
the corresponding contributions due to the exchange by pseudoscalar and scalar
mesons. The basis of our model calculations is the spontaneous violation of
the chiral symmetry in the model with the nonlocal four-fermion interaction
and abelian gauge invariance. The first leads to the generation of the
momentum-dependent dynamical quark mass, and the latter ensures the
fulfillment of the Ward-Takahashi identities with respect to the quark-photon interaction.

In the present work, we derived the general
expression for $a_{\mu}^{\mathrm{LbL}}$ as the three-dimensional integral in
modulus of the two photon momenta and the angle between them. The integral is
the convolution of the known kinematical factors and some projections of the
four-photon polarization tensor. The latter is the subject of theoretical calculations.

Since our model calculations of the hadronic contributions are basically
numerical, it is more convenient to present our results in terms of the
density function $\rho^{\mathrm{HLbL}}(Q_{1},Q_{2})$. We observe some
properties of this function that have model-independent character. Firstly, at
zero momenta one has \\$\rho^{\mathrm{HLbL}}(0,0)=0$ in spite of the fact that
the partial contributions of different diagrams are nonzero in this limit.
This low-energy theorem is a direct consequence of the quark-photon gauge
invariance and the spontaneous violation of the chiral symmetry. Secondly, at
high momenta the density is saturated by the contribution from the box diagram
with the local quark-photon vertices and local quark propagators in accordance
with the perturbative theory. This is a consequence of the fact, that at small
distances all nonperturbative nonlocal effects are washed out. Thirdly, with
the model parameters chosen, the $\rho^{\mathrm{HLbL}}(Q_{1},Q_{2})$ is
concentrated in the region $Q_{1}\approx Q_{2}\approx300$ MeV, which is a
typical scale for light hadrons.

Summarizing the results of the present and previous works
\cite{Dorokhov:2011zf,Dorokhov:2012qa,Dorokhov:2014iva}, we get the total
hadronic contribution to $a_{\mu}^{\mathrm{HLbL}}$ within the N$\chi$QM in the
leading order in the $1/N_{c}$ expansion. The total result is given in Eq.
(\ref{aNchiQM}). To estimate the uncertainty of this result, we vary some of
the model parameters in physically reasonable interval and also study the
sensitivity of the result with respect to different model parameterizations.
In this sense, the error in Eq. (\ref{aNchiQM}) is a conservative one.

If we add the result (\ref{aNchiQM}) to all other known contributions of the
standard model to $a_{\mu}$, (\ref{aQED})-(\ref{aNNLO}),
we get
that the difference between experiment (\ref{amuBNL2}) and theory
is
\begin{equation}
a_{\mu}^{\mathrm{BNL,CODATA}}-a_{\mu}^{\mathrm{SM}}=18.73\times10^{-10},
\end{equation}
which corresponds to $2.43\sigma$. If one uses the hadronic vacuum
polarization contribution from the $\tau$ hadronic decays instead of
$e^{+}e^{-}$ data
\begin{equation}
a_{\mu}^{\mathrm{HVP,LO-\tau}}=701.5(4.7)\times10^{-10}\qquad
\text{\cite{Davier:2010nc}},
\end{equation}
the difference decreases to $18.44\times10^{-10}$ ($2.23\sigma$) for the case
of $a_{\mu}^{\mathrm{HLbL}}$ from (\ref{aHLbLprades}) \cite{Prades:2009tw} and
to $12.14\times10^{-10}$ ($1.53\sigma$) in our model (\ref{aNchiQM}).

Clearly, a further reduction of both the experimental and theoretical
uncertainties is necessary. On the theoretical side, the calculation of the
still badly known hadronic light-by-light contributions in the next-to-leading
order in the $1/N_{c}$ expansion 
(the pion and kaon loops) and extension of the model by including heavier vector and axial-vector mesons 
is the next goal.
The contribution of these effects and the model error induced by them are not included in the result (\ref{aNchiQM}).
Preliminary studies \cite{
Hayakawa:1995ps,Melnikov:2003xd} show that these contributions
are one order smaller than the pseudoscalar exchanges and the quark loop contributions.
However, the interesting point that inclusion of vector
channel can strongly suppress contribution from the quark loop
due photon--vector meson exchange which lead
to appearance in each photon vertex additional VMD-like factor.
This was found in local NJL model \cite{Bijnens:1995cc} and should be
carefully investigated in the nonlocal one.

Work in this direction is
now in progress, and we hope to report its results in the near future.
\begin{acknowledgements}
We thank J. Bijnens, Yu.M. Bystritskiy, A.L. Kataev, N.I. Kochelev, \fbox{E.A. Kuraev},
V.P. Lomov, A.~Nyffeler, H.-P. Pavel, and A.A. Pivovarov for critical remarks
and illuminating discussions. Numerical calculations are performed on
computing cluster "Ac. V.M. Matrosov". The work is supported by Russian Science Foundation grant (RSCF 15-12-10009).
\end{acknowledgements}
\appendix
\section{Nonlocal multi-photon vertices}
Let us introduce the finite-difference derivatives%
\begin{align}
&f^{\left(  1\right)  }\left(  a,b\right)     =\frac{f\left(  a+b\right)
-f\left(  b\right)  }{\left(  a+b\right)  ^{2}-b^{2}},\label{FDD1}\\
&f^{(n+1)}\left(  a,\{b_{i}\},b_{1},b_{2}\right)     =\frac{f^{(n)}\left(
a,\{b_{i}\},b_{1}\right)  -f^{(n)}\left(  a,\{b_{i}\},b_{2}\right)  }{\left(
a+b_{1}\right)  ^{2}-\left(  a+b_{2}\right)  ^{2}},\nonumber
\end{align}
where $n=1,2,...$ .
Then, the quark-antiquark
vertex with the two-photon insertions (Fig. \ref{fig:VerticesPhot-n}b) is \cite{Terning:1991yt}
\begin{align}
&  \Gamma_{\mu\nu}^{\left(  2\right)  }\left(  q_{1},q_{2}\right)  =2g_{\mu
\nu}m^{\left(  1\right)  }\left(  k,k^{\prime}\right)  +\nonumber\\
&  \quad\left(  k+k_{1}\right)  _{\mu}\left(  k_{1}+k^{\prime}\right)  _{\nu
}m^{\left(  2\right)  }\left(  k,k_{1},k^{\prime}\right)  +\label{DGam2}\\
&  \quad\left(  k+k_{2}\right)  _{\nu}\left(  k_{2}+k^{\prime}\right)  _{\mu
}m^{\left(  2\right)  }\left(  k,k_{2},k^{\prime}\right)  .\nonumber
\end{align}
Here and below, $k$ is the momentum of the incoming quark, $k^{\prime}$ is the
momentum of the outgoing quark, $q_{i}$ are the momenta of the incoming
photons, and $k_{1}=k+q_{1},$ $k_{ij...k}=k+q_{i}+q_{j}+...+q_{k}$.

The quark-three-photon vertex (Fig. \ref{fig:VerticesPhot-n}c) is \cite{Liao:2008dd}
\begin{align}
& \Gamma_{\mu\nu\rho}^{(3)}(q_{1},q_{2},q_{3})   =-\big[
2g_{\mu\nu}
(k_{12}+k^{\prime})_{\rho}m^{(2)}(k,k_{12},k^{\prime})\nonumber\\
&\quad +2g_{\mu\nu}(k+k_{3})_{\rho}%
m^{(2)}(k,k_{3},k^{\prime})
\nonumber\\
&\quad  +(k+k_{1})_{\mu}(k_{1}+k_{12})_{\nu}(k_{12}+k^{\prime})_{\rho}%
m^{(3)}(k,k_{1},k_{12},k^{\prime})\nonumber\\
&\quad  +(k+k_{1})_{\mu}(k_{13}+k^{\prime})_{\nu}(k_{1}+k_{13})_{\rho}%
m^{(3)}(k,k_{1},k_{13},k^{\prime})\big]\nonumber\\
&\quad  +[1\rightleftarrows3,\mu\rightleftarrows\rho]+[2\rightleftarrows
3,\nu\rightleftarrows\rho].\label{Gammappqqq}
\end{align}
\allowdisplaybreaks
The quark-four-photon vertex (Fig. \ref{fig:VerticesPhot-n}d) takes the form
\begin{align}
&\Gamma_{\mu\nu\rho\tau}^{(4)}(q_{1},q_{2},q_{3},q_{4})=\Big[
+4g_{\mu\nu}g_{\tau\rho}m^{(2)}(k,k_{12},k^{\prime}) \nonumber\\
&+4g_{\mu\nu}g_{\tau\rho}m^{(2)}(k,k_{34},k^{\prime})\nonumber\\
&+2g_{\mu\nu}\Big(    (k_{{}}+k_{3})_{\rho}(k_{3}+k_{34})_{\tau}%
m^{(3)}(k,k_{3},k_{34},k^{\prime})\nonumber\\
&  +(k_{{}}+k_{3})_{\rho}(k_{123}+k^{\prime})_{\tau}m^{(3)}(k,k_{3}%
,k_{123},k^{\prime})\nonumber\\
&  +(k_{12}+k_{123})_{\rho}(k_{123}+k^{\prime})_{\tau}m^{(3)}(k,k_{12}%
,k_{123},k^{\prime})\nonumber\\
&  +(k_{124}+k^{\prime})_{\rho}(k_{12}+k_{124})_{\tau}m^{(3)}(k,k_{12}%
,k_{124},k^{\prime})\nonumber\\
&  +(k_{124}+k^{\prime})_{\rho}(k_{{}}+k_{4})_{\tau}m^{(3)}(k,k_{4}%
,k_{124},k^{\prime})\nonumber\\
&  +(k_{4}+k_{34})_{\rho}(k_{{}}+k_{4})_{\tau}m^{(3)}(k,k_{4},k_{34}%
,k^{\prime})\Big)\label{Gqqqq4}\\
&+2g_{\tau\rho}\Big(    (k_{{}}+k_{1})_{\mu}(k_{1}+k_{12})_{\nu}%
m^{(3)}(k,k_{1},k_{12},k^{\prime})\nonumber\\
&  +(k_{{}}+k_{2})_{\nu}(k_{2}+k_{12})_{\mu}m^{(3)}(k,k_{2},k_{12},k^{\prime
})\nonumber\\
&  +(k_{34}+k_{234})_{\nu}(k_{234}+k_{1234})_{\mu}m^{(3)}(k,k_{34}%
,k_{234},k^{\prime})\nonumber\\
&  +(k_{34}+k_{134})_{\mu}(k_{134}+k_{1234})_{\nu}m^{(3)}(k,k_{34}%
,k_{134},k^{\prime})\nonumber\\
&  +(k_{{}}+k_{1})_{\mu}(k_{134}+k_{1234})_{\nu}m^{(3)}(k,k_{1},k_{134}%
,k^{\prime})\nonumber\\
&  +(k_{{}}+k_{2})_{\nu}(k_{234}+k_{1234})_{\mu}m^{(3)}(k,k_{2},k_{234}%
,k^{\prime})\Big)\nonumber\\
&  +(k_{{}}+k_{1})_{\mu}(k_{1}+k_{12})_{\nu}(k_{12}+k_{123})_{\rho}%
(k_{123}+k^{\prime})_{\tau}\times \nonumber \\&\qquad\qquad\times m^{(4)}(k,k_{1},k_{12},k_{123},k^{\prime
})\nonumber\\
&  +(k_{{}}+k_{1})_{\mu}(k_{1}+k_{12})_{\nu}(k_{124}+k_{1234})_{\rho}%
(k_{12}+k_{124})_{\tau}\times \nonumber \\&\qquad\qquad\times m^{(4)}(k,k_{1},k_{12},k_{124},k^{\prime})\nonumber\\
&  +(k_{2}+k_{12})_{\mu}(k_{{}}+k_{2})_{\nu}(k_{12}+k_{123})_{\rho}%
(k_{123}+k_{1234})_{\tau}\times \nonumber \\&\qquad\qquad\times m^{(4)}(k,k_{2},k_{12},k_{123},k^{\prime})\nonumber\\
&  +(k_{2}+k_{12})_{\mu}(k_{{}}+k_{2})_{\nu}(k_{124}+k_{1234})_{\rho}%
(k_{12}+k_{124})_{\tau}\times \nonumber \\&\qquad\qquad\times m^{(4)}(k,k_{2},k_{12},k_{124},k^{\prime})\nonumber\\
&  +(k_{23}+k_{123})_{\mu}(k_{{}}+k_{2})_{\nu}(k_{2}+k_{23})_{\rho}%
(k_{123}+k_{1234})_{\tau}\times \nonumber \\&\qquad\qquad\times m^{(4)}(k,k_{2},k_{23},k_{123},k^{\prime})\nonumber\\
&  +(k_{24}+k_{124})_{\mu}(k_{{}}+k_{2})_{\nu}(k_{124}+k_{1234})_{\rho}%
(k_{2}+k_{24})_{\tau}\times \nonumber \\&\qquad\qquad\times m^{(4)}(k,k_{2},k_{24},k_{124},k^{\prime})\nonumber\\
&  +(k_{234}+k_{1234})_{\mu}(k_{{}}+k_{2})_{\nu}(k_{24}+k_{234})_{\rho}%
(k_{2}+k_{24})_{\tau}\times \nonumber \\&\qquad\qquad\times m^{(4)}(k,k_{2},k_{24},k_{234},k^{\prime})\nonumber\\
&  +(k_{234}+k_{1234})_{\mu}(k_{{}}+k_{2})_{\nu}(k_{2}+k_{23})_{\rho}%
(k_{23}+k_{234})_{\tau}\times \nonumber \\&\qquad\qquad\times m^{(4)}(k,k_{2},k_{23},k_{234},k^{\prime}%
)\Big]+\nonumber\\
&  +[2\leftrightarrows4,\nu\leftrightarrows\tau]+[2\leftrightarrows
3,\nu\leftrightarrows\rho].\nonumber
\end{align}

\begin{table*}[thb]
\label{table1}
\centering
\begin{tabular}{|c| c c c c c|c|c|c|c|c|c|}
\hline
\multirow{3}*{No.}&\multicolumn{5}{c|}{model parameters}&\multicolumn{6}{c|}{$a_{\protect\mu }^{\mathrm{LbL}}$ in $10^{-10}$ }\\\cline{2-6}\cline{7-12}
               & $m_{d,u}$ & $m_{c,u}$ & $m_{d,s}$ & $m_{c,s}$ & $\Lambda$    & total    &  total   & $u+d$       & $s$ quark & \multirow{2}*{$\pi^0+\sigma$} & $\eta+\eta^{\prime}+$ \\
               & MeV       & MeV       & MeV       & MeV       &MeV           & $ SU(3)$ & $ SU(2)$ &  quark loop &  loop     &   & $a_0+f_0$ \\
\hline
$1$  & $150$   & $0.33$ &         &         & $6786.6$ &         & $21.74$ & $16.95$ &         & $4.79$ &        \\
$2$  & $160$   & $0.59$ &         &         & $4890.7$ &         & $20.49$ & $15.38$ &         & $5.11$ &        \\
$3$  & $170$   & $0.93$ &         &         & $3768.9$ &         & $19.45$ & $14.10$ &         & $5.35$ &        \\
$4$  & $180$   & $1.33$ &         &         & $3049.2$ &         & $18.60$ & $13.10$ &         & $5.50$ &        \\
$5$  & $190$   & $1.78$ &         &         & $2557.6$ &         & $17.89$ & $12.29$ &         & $5.60$ &        \\
$6$  & $200$   & $2.27$ &         &         & $2204.9$ &         & $17.30$ & $11.67$ &         & $5.64$ &        \\
$7$  & $210$   & $2.79$ &         &         & $1941.5$ &         & $16.82$ & $11.18$ &         & $5.64$ &        \\
$8$  & $220$   & $3.34$ &         &         & $1738.2$ &         & $16.42$ & $10.80$ &         & $5.62$ &        \\
$9$  & $230$   & $3.90$ &         &         & $1577.1$ &         & $16.09$ & $10.52$ &         & $5.58$ &        \\
$10$ & $240$   & $4.47$ &         &         & $1446.3$ &         & $15.83$ & $10.31$ &         & $5.52$ &        \\
$11$ & $240$   & $4.47$ & $339.5$ & $133.7$ & $1446.3$ & $16.53$ & $15.89$ & $10.31$ & $0.234$ & $5.58$ & $0.404$\\
$12$ & $250$   & $5.06$ &         &         & $1338.2$ &         & $15.61$ & $10.16$ &         & $5.46$ &        \\
$13$ & $250$   & $5.06$ & $347.1$ & $148.4$ & $1338.2$ & $16.21$ & $15.64$ & $10.16$ & $0.229$ & $5.49$ & $0.337$\\
$14$ & $260$   & $5.65$ &         &         & $1247.2$ &         & $15.45$ & $10.05$ &         & $5.39$ &        \\
$15$ & $270$   & $6.25$ &         &         & $1169.6$ &         & $15.31$ & $9.99$  &         & $5.33$ &        \\
$16$ & $280$   & $6.86$ &         &         & $1102.5$ &         & $15.23$ & $9.97$  &         & $5.26$ &        \\
$17$ & $280$   & $6.86$ & $387.4$ & $193.4$ & $1102.5$ & $15.91$ & $15.22$ & $9.97$  & $0.209$ & $5.25$ & $0.480$\\
$18$ & $290$   & $7.48$ &         &         & $1043.9$ &         & $15.14$ & $9.94$  &         & $5.19$ &        \\
$19$ & $300$   & $8.09$ &         &         & $992.2$  &         & $15.08$ & $9.96$  &         & $5.12$ &        \\
$20$ & $305$   & $8.41$ & $413.7$ & $231.9$ & $968.6$  & $15.78$ & $15.05$ & $9.97$  & $0.193$ & $5.08$ & $0.543$\\
$21$ & $310$   & $8.72$ &         &         & $946.2$  &         & $15.04$ & $9.98$  &         & $5.06$ &        \\
$22$ & $320$   & $9.35$ &         &         & $904.9$  &         & $15.02$ & $10.02$ &         & $5.00$ &        \\
$23$ & $320$   & $9.35$ & $428$   & $255.7$ & $904.9$  & $15.78$ & $15.02$ & $10.03$ & $0.182$ & $4.99$ & $0.577$\\
$24$ & $330$   & $9.99$ &         &         & $867.7$  &         & $15.01$ & $10.08$ &         & $4.94$ &        \\
$25$ & $340$   & $10.63$&         &         & $833.8$  &         & $15.02$ & $10.14$ &         & $4.88$ &        \\
$26$ & $350$   & $11.29$&         &         & $802.8$  &         & $15.03$ & $10.21$ &         & $4.82$ &        \\
$27$ & $350$   & $11.29$& $451.2$ & $305.9$ & $802.8$  & $15.80$ & $15.05$ & $10.21$ & $0.159$ & $4.84$ & $0.592$\\
$28$ & $360$   & $11.95$&         &         & $774.4$  &         & $15.07$ & $10.30$ &         & $4.77$ &        \\
$29$ & $370$   & $12.62$&         &         & $748.1$  &         & $15.10$ & $10.38$ &         & $4.72$ &        \\
$30$ & $380$   & $13.30$&         &         & $723.8$  &         & $15.14$ & $10.47$ &         & $4.67$ &        \\
$31$ & $390$   & $13.99$&         &         & $701.1$  &         & $15.19$ & $10.56$ &         & $4.63$ &        \\
$32$ & $400$   & $14.69$&         &         & $679.8$  &         & $15.24$ & $10.65$ &         & $4.58$ &        \\
\hline
$\mathrm{G}_{I}$   & $304.5$ & $8.50$ & $427$   & $223$   & $1002.7$ & $15.67$ & $14.67$ & $9.71$  & $0.192$ & $4.95$ & $0.810$\\
$\mathrm{G}_{II}$  & $304.5$ & $8.50$ & $439$   & $223$   & $1002.7$ & $15.93$ & $14.67$ & $9.71$  & $0.190$ & $4.95$ & $1.070$\\
$\mathrm{G}_{III}$ & $304.5$ & $8.50$ & $422$   & $223$   & $1002.7$ & $15.57$ & $14.67$ & $9.71$  & $0.193$ & $4.95$ & $0.707$\\
$\mathrm{G}_{IV}$  & $287.5$ & $7.50$ & $408$   & $199$   & $1086.1$ & $15.75$ & $14.81$ & $9.75$  & $0.202$ & $5.06$ & $0.738$\\
\hline
$\mathrm{L}_{I}$   & $295$   & $4.00$ & $450$   & $112$   & $1013$   & $15.58$ & $14.84$ & $9.61$  & $0.241$ & $5.23$ & $0.503$\\
$\mathrm{L}_{II}$  & $295$   & $4.00$ & $505$   & $110$   & $1013$   & $16.37$ & $14.84$ & $9.61$  & $0.222$ & $5.23$ & $1.311$\\
$\mathrm{L}_{III}$ & $296$   & $4.00$ & $457$   & $112$   & $1013$   & $15.61$ & $14.80$ & $9.57$  & $0.238$ & $5.22$ & $0.578$\\
$\mathrm{L}_{IV}$  & $277.5$ & $3.50$ & $418$   & $100$   & $1110$   & $15.76$ & $15.08$ & $9.74$  & $0.251$ & $5.33$ & $0.435$\\
\hline
\end{tabular}%
\caption{The contribution to the muon
AMM $a_{\protect\mu }^{\mathrm{LbL}}$ for different
sets of model parameters. The model parameters $\mathrm{G}_{I-IV}$ and $\mathrm{L}_{I-IV}$ are taken from
 \cite{Scarpettini:2003fj}.
The difference between $a_{\protect\mu }^{\mathrm{LbL}}$ for the set of model parameters with same $u$ quark mass
in $SU(2)$ and $SU(3)$ models is due to mixing of $\sigma$ meson with $f_0$ meson is $SU(3)$ case.
In order to extrapolate the $SU(3)$-result to lower
quark masses, we find that the maximal value of the difference between the
total $SU(2)$- and $SU(3)$-results is $0.77$. We add this number to the value
of the $SU(2)$-result at $200$ MeV.
}
\end{table*}
\end{document}